\documentclass[aps,prev,twocolumn,preprintnumbers,floatfix,nofootinbib]{revtex4-1}
\usepackage[english]{babel}
\usepackage{bm}
\usepackage{times}
\usepackage{ulem}
\usepackage{listings}
\usepackage{slashed}
\usepackage{color}
\usepackage{appendix}
\usepackage{mathtools}
\usepackage{natbib}
\usepackage{epsfig}
\usepackage{dcolumn}
\usepackage{textcomp}
\usepackage{multirow}
\usepackage{graphicx}
\usepackage{soul}
\usepackage{subfigure}
\usepackage{xcolor}
\usepackage[capitalize]{cleveref}

\begin{document}

%\title{Updated BBN Bounds on Energy Injection in the Early Universe}
\title{Updated Big Bang Nucleosynthesis Bounds on Long-lived Particles from Dark Sectors}

\author{J. R. Alves$^{1,2}$}
\email{jose.mourafisica99@gmail.com}
\author{L. Angel$^{1,2}$}
\email[Corresponding author:]{silucoan@gmail.com}
\author{L. Guedes$^{1,2}$}
\email[Corresponding author:]{leticiamvguedes@gmail.com}
\author{R. M. P. Neves$^{2}$}
\email{raissapimentel.ns@gmail.com}
\author{F. S. Queiroz$^{1,2,3}$}
\email{farinaldo.queiroz@ufrn.br}
\author{D. R. da Silva$^{2,5}$}
\email{deivid.rodrigo.ds@gmail.com}
\author{R. Silva$^{1}$}
\email{raimundosilva@fisica.ufrn.br}
\author{Y. Villamizar$^{6}$}
\email{yoxarasv@gmail.com}

\affiliation{$^{1}$Departamento de F\'isica, Universidade Federal do Rio Grande do Norte, 59078-970, Natal, RN, Brasil}
\affiliation{$^2$International Institute of Physics, Universidade Federal do Rio Grande do Norte,
Campus Universitario, Lagoa Nova, Natal-RN 59078-970, Brazil}
\affiliation{$^3$Millennium Institute for Subatomic Physics at the High-Energy Frontier (SAPHIR) of ANID, Fern\'andez Concha 700, Santiago, Chile}
\affiliation{$^4$Departamento de Fisica, Universidade Federal da Paraiba, 58051-970, Jo\~ao Pessoa, PB, Brasil}
\affiliation{$^5$Centro Brasileiro de Pesquisas F\'{\i}sicas, Rua Dr. Xavier Sicaud 150, Urca, CEP-22290-180, Rio de Janeiro, RJ, Brasil}
\affiliation{$^6$ Centro de Ci\^encias Naturais e Humanas, Universidade Federal do ABC, 09210-580, Santo Andr\'e, S\~ao Paulo, Brasil}

\begin{abstract}
\noindent As electromagnetic showers may alter the abundance of Helium, Lithium, and Deuterium, we can place severe constraints on the lifetime and amount of electromagnetic energy injected by long-lived particles. Considering up-to-date measurements of the light element abundances that point to $Y_p=0.245\pm 0.003$, $({\rm D/H})= (2.527\pm 0.03)\times 10^{-5}$, and the baryon-to-photon ratio obtained from the Cosmic Microwave Background data, $\eta=6.104 \times 10^{-10}$, we derive upper limits on the fraction of electromagnetic energy produced by long-lived particles. Our findings apply to decaying dark matter models, long-lived gravitinos, and other non-thermal processes that occurred in the early universe between $10^2-10^{10}$ seconds. 
\end{abstract}

\maketitle

\section{\label{Intro}Introduction}

There are four pillars that validate the hot Big Bang Theory, namely the Planck spectrum of the Cosmic Microwave Background (CMB), the density fluctuations imprinted in the CMB  and the distribution of galaxies, the confirmed expansion of the universe, and the Big Bang Nucleosynthesis (BBN). Collectively, these observations support the notion that the universe originated from a state of high temperature and density. BBN is rooted in thoroughly studied physics processes, encompassing the spectra, reactions, and weak decays of light elements. The BBN prediction for the light elements abundances such as Deuterium, Helium-3, and Helium-4 and Lithium-7 rely solely on one parameter, the baryon-photon ratio, $\eta=n_b/n_\gamma$. The ratio of any two primordial abundances should give $\eta$, and the measurement of the other three elements tests the BBN theory. The abundances of the light elements have been measured in several terrestrial and astrophysical environments. Although, it has often been hard to determine when these abundances are close to the primordial ones. Nevertheless, there is a clear agreement with the BBN predictions for all the light nuclei. The significant improvement experienced in the measurements of the light element abundances has deepened our knowledge of the primordial universe. For instance, for a long time the error on the baryon density, $\Omega_b$ was large enough to allow a dark baryonic component in the universe. In the late 90's the error was too large, $\Omega_b h^2=0.009-0.02$ \cite{Copi:1994ev}. New measurements of the Deuterium abundance, D/H, using quasars observations point to $\Omega_b h^2= 0.022$, with $h=0.674$ \cite{ParticleDataGroup:2022pth}, solidly confirming the existence of a dark matter component in the universe, which is essentially non-baryonic. In other words, BBN stands as evidence for dark matter in the universe.  We highlight that Planck data on the CMB point to $\Omega_b h^2 = 0.02230 \pm 0.00021$ and $\eta=6.104 \pm 0.058 \times 10^{-10}$ \cite{Planck:2018vyg,Fields:2019pfx}, confirming BBN predictions. 

 Interestingly, BBN is a phenomenon that unfolds within a well-defined sequence of events during the first minutes of the Universe. The remarkable agreement between BBN predictions and observational data establishes BBN as an important probe for new physics events that might alter the abundance of those elements, which is a fact that will be explored in this work. In particular, BBN places stringent constraints on new physics beyond the Standard Model that injects energy into the cosmological plasma, such as the decays of massive particles with lifetimes greater than $0.1$~s \cite{Forestell:2018txr}. 

 There have been many works that discussed in great detail the impact of energy injection in the BBN, and used this information to derive limits on the lifetime of long-lived massive particles whose decays induce electromagnetic or hadronic interactions with the plasma, leading to processes that alter the abundance of the light elements. Hadrons produced from the decay of a long-lived particle scatter off protons and neutrons modifying the ratio of these baryons, and consequently the helium abundance. They can also destroy and modify the abundance of helium and other light elements through hadrodissociation. The chain reactions are non-trivial and the final abundance of the elements is highly sensitive to the energy injected in the form of hadrons into the plasma \cite{Forestell:2018txr}.

Supersymmetric models \cite{Feng:2003uy,Feng:2004zu,Allahverdi:2014bva}, and enlarged gauge groups \cite{Kelso:2013nwa}, and others \cite{Hooper:2011aj,Hufnagel:2018bjp,Depta:2019lbe} feature particles with a long lifetime, $\tau > 10^2$~s. When long-lived particles inject electromagnetic energy into the plasma, i.e., photons, electrons, and positrons, the abundance of light elements is only meaningfully affected when the decay happens at later times, $\tau > 10^4$~s. Photodissociation plays a major role in the abundance of light elements compared to other processes. In this work, we do not aim to go into detail about the physical processes and review all technical computations, for this matter, we refer the reader to \cite{Tytler:2000qf}. We will rather present updated limits on the lifetime and the amount of electromagnetic energy that long-lived particles could inject into the early universe using the most updated values of the light elements abundances. Previous works placed limits on long-lived particles using the abundance of D, $^4$He, and $^7$Li only. Given the significant uncertainties surrounding the primordial abundance of $^6$Li and  $^7$Li, we have chosen not to incorporate those abundances into our parameter space constraints.

In summary, our work is structured as follows: In section \ref{secII} we briefly review the primordial abundances of the elements; in section \ref{secIII} we discuss the key ingredients for computing the electromagnetic cascade resulting from the decay of the long-lived particles; in section \ref{secIV} we present our updated limits; in section \ref{secV} we draw our conclusions.

%https://arxiv.org/pdf/1011.1054.pdf
%https://arxiv.org/pdf/1809.01179.pdf
%https://arxiv.org/pdf/astro-ph/0001318.pdf
%https://pdg.lbl.gov/2022/reviews/rpp2022-rev-bbang-nucleosynthesis.pdf

\section{Abundance of the Light Elements \label{secII}}

The abundance of the light elements is determined by a set of first-order differential Boltzmann equations originally developed in \cite{Alpher:1948ve,Gamow:1948pob,Hayashi:1950lqo},
\begin{eqnarray}
\label{start1}
\frac{dY_i}{dt}=-H(T)T \frac{d Y_i}{ dT } = \sum (\Gamma_{ij}Y_j+ \Gamma_{ikl}  Y_k Y_l+...),
\end{eqnarray}where $Y_i = n_i / n_b$ is the ratio between
the number density of the species $i$ and baryon number density. The relevant interaction and decay rates are encoded in
$\Gamma_{ij}$ and $\Gamma_{ikl}$, respectively. $H(T)$ is the hubble rate given by, 
\begin{equation}
\label{Hubbleeq}
H(T) = T^2 \left(\frac{8\pi^3 g_* G_N}{90}\right)^{1/2}
\end{equation} with $g_* = g_{bosons} +\frac{7}{8}g_{fermions}$ corresponding to the relativistic degrees of freedom
of the species at the time, $G_N$ being the gravitation constant, and T the plasma temperature. Eq.\ref{start1} must be solved considering that after electron-positron annihilation, neutrinos, and photons have different temperatures, $T_\nu \simeq (4/11)^{1/3} T_\gamma$.

Several well-known processes occur and dictate the production of light elements.  We briefly review them below.\\

\paragraph{Helium-4}

Weak interactions govern the n-p conversion, and they scale with $\Gamma_{n-p} \sim G_F^2 T^5$, where $G_F$ is the Fermi
constant. The universe cools down with $H(T) \sim \sqrt{g_{\ast}G_N}T^2$, and eventually, the interaction rate fell faster than the Hubble rate, resulting in a departure from chemical equilibrium (freeze-out) around $T\sim 1$~MeV. At this temperature, the ratio between the number of neutrons to proton is $n/p \sim \exp^{m_n-m_p/1MeV}\sim 1/6$, but continues to decrease down to $1/7$ due to $\beta$ decay. 

All neutrons form  Helium-4 at this time, because many background photons delay deuterium formation through photon-dissociation. Hence, the $^4$He mass fraction $Y_p$ is found to be \cite{Pospelov:2010hj},
\begin{equation}
  Y_p \simeq \frac{2n/p}{1+n/p}=0.248
\end{equation}

Therefore, $^4$He mass fraction is amenable to the neutron-to-proton freeze-out and timing at which Deuterium production becomes efficient.  

\paragraph{Deuterium}

\begin{figure}[ht!]
\centering
    \includegraphics[width =\columnwidth]{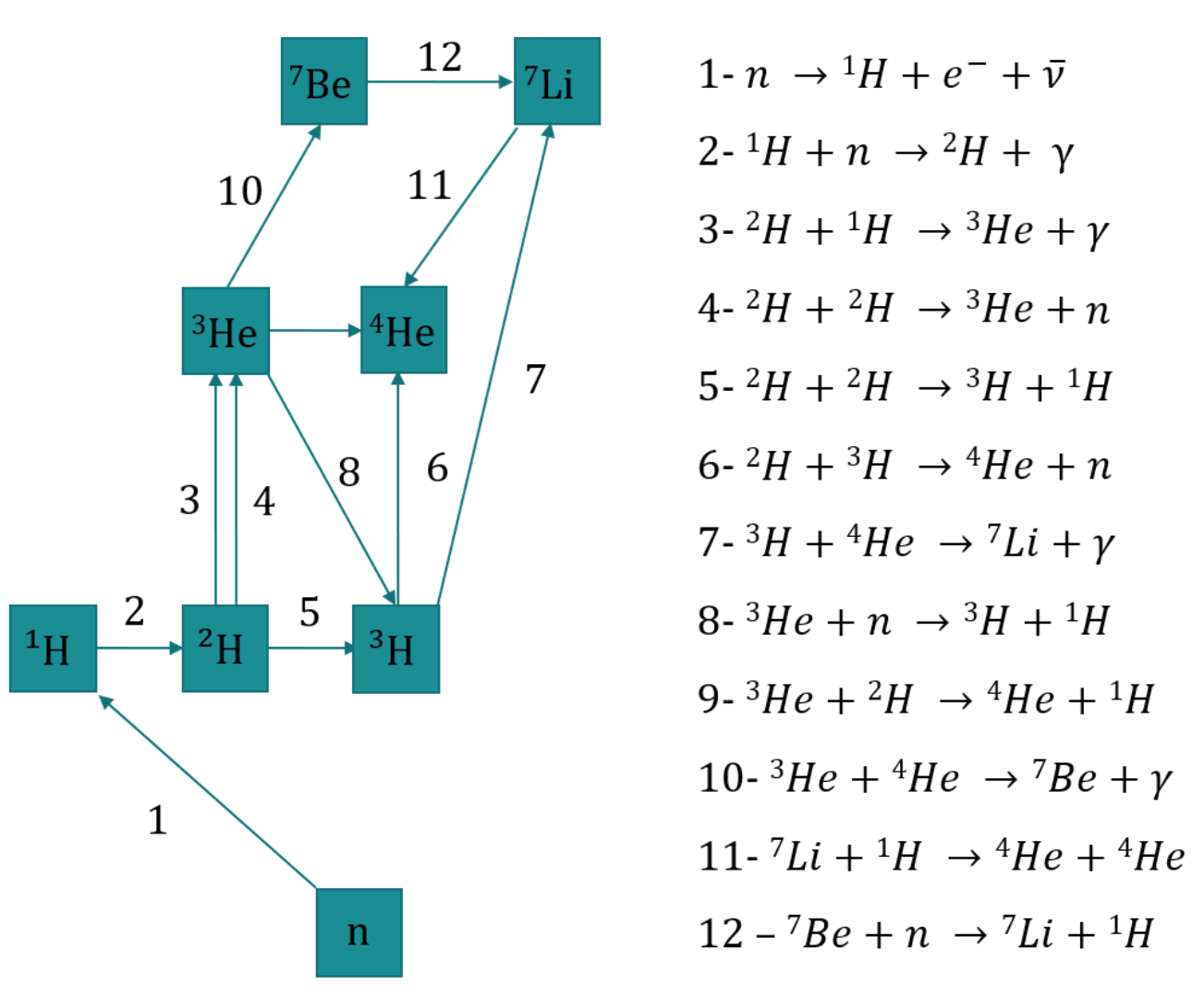}
     \caption{ \label{fig:BBNchain} Representation of the relevant chain reactions for BBN.}
\end{figure}

When the temperature fall below $70$~keV, the exponential Boltzmann suppression on the number of photons is sufficient to significantly induce the production of Deuterium (D) and consequently initiate other nuclear reactions (See the illustration in Fig.\ref{fig:BBNchain}).  The Deuterium and  $^3$He productions occur concurrently. Most of the deuterium then collided with other protons and neutrons to produce helium and a small amount of tritium. Deuterium also plays a minor role in the production of Lithium 7.

\paragraph{Lithium-7}
The rate for producing elements with A=6,7 is smaller than the Hubble rate. Consequently, the abundances are tiny. As far as $^7$Li is concerned,  around 90\% of the primordial $^7$Li stems $^7$Be having in mind the $\eta$ value inferred from the CMB.

\paragraph{Lithium-6}
The primordial abundance of $^6$Li is often not used to constrain new physics due to the uncertainties involved. Similarly to $^7$Li, the abundance of $^6$Li is also measured in halo stars \cite{Asplund:2005yt}. Observations lead to a $^6$Li/$^7$Li ratio of $\sim 0.05$. One could use the $^6$Li abundance to set constraints on new physics. However, due to the non-trivial assessment of its primordial abundance, we will not consider it in our results.

After this brief review of the light element abundances, we will address the electromagnetic injection.

\section{\label{secIII}Electromagnetic Injection}

New physics processes that inject electrons, positrons, or photons can induce photodissociation reactions, which is the most important effect on the light elements abundances. These particles lose energy very efficiently by scattering off the highly abundant photon background. The scattering process has an energy threshold. If the energy of the injected photon is sufficiently large ($E_\gamma> m_e^2/E_{\gamma,BG}$), typically the pair production $\gamma +\gamma \rightarrow e^+ e^-$ dominates because both cross-section and number density of the background photon are large \cite{Protheroe:1994dt,Kawasaki:1994sc}.  Below this energy threshold,  photons lose energy mostly via photon-photon scattering $(\gamma +\gamma_{BG} \rightarrow \gamma +\gamma)$, Compton scattering ($\gamma +e^{\pm} \rightarrow \gamma e^{\pm}$) and pair creation on nuclei ($\gamma + N_{BG} \rightarrow e^+ e^- N_{BG}$). However, if the new physics process injects high-energy electrons or positrons, then inverse Compton scattering governs the energy loss \cite{Kawasaki:1994sc}. Comparing the interaction rate with the expansion rate, we conclude that these processes are not very efficient at early times, $\tau < 10^4$~s. Therefore, these energy injection episodes are seen as a post-event that can alter the BBN predictions. 

In summary, one can account for these possibilities by including the non-thermal photon population and abundances of nuclei from,
\begin{eqnarray}
  -HT\frac{dY_B}{dT} =  \sum_{A} Y_A \int_0^\infty dE_\gamma f_{\gamma}^{\rm qse}(E_\gamma)
  \sigma_{\gamma +A \rightarrow B}(E_\gamma) \nonumber \\ 
  - Y_B \sum_{C} \int_0^\infty dE_\gamma f_{\gamma}^{\rm qse}(E_\gamma)
  \sigma_{\gamma +B \rightarrow C}(E_\gamma).
\end{eqnarray} where the energy spectrum is found to be ~\cite{Cyburt:2002uv},
\begin{equation}
f_{\gamma}^{\rm qse} = \frac{n_X p_\gamma(E_\gamma)}{\tau_X \Gamma_\gamma(E_\gamma)},
\end{equation}with $n_X$ being the number density of the long-lived particle, and $\tau_X$ its lifetime, and $p_\gamma$ the photon spectrum which we assumed to be approximated by a power law~\cite{Protheroe:1994dt},
\begin{align}
  p_{\gamma}(E_{\gamma}) = 
\begin{cases} 
K_0 (E_{\gamma}/E_{\mathrm{low}})^{-1.5}  &\mathrm{for}\
  E_{\gamma}<E_{\mathrm{low}} \\%[-0.3cm]
 K_0 (E_{\gamma}/E_{\mathrm{low}})^{-2.0} & \mathrm{for}\
  E_{\mathrm{low}}<E_{\gamma}<E_C \\%[-0.3cm]
\qquad\quad 0 & \mathrm{for}\ E_{\gamma}>E_C
\end{cases}
\label{eq:pgamma}
\end{align}where $E_{\rm low} \simeq m_e^2/(80T)$, and the energy threshold of pair production $E_C
\simeq m_e^2/22T$~\cite{Kawasaki:1994sc}. The normalization constant $K_0$ in Eq.\ref{eq:pgamma} is obtained by requiring that the electromagnetic injected is given by $E_X = \int d E_\gamma E_\gamma p_\gamma$. We emphasize that the power law approximation is a good approximation for energies larger than $E_C$ \cite{Poulin:2015opa,Forestell:2018txr} as we consider here. That said, we can compute the change in the abundance of the light elements as a function of the lifetime $\tau$ and the total electromagnetic energy released $\zeta_X$, where $\tau$ is the lifetime of the decaying particle. The total electromagnetic energy released in a process of the type $X^\prime \rightarrow X +\gamma$ as we will consider here is given by,

\begin{equation}
\zeta_X  =E_\gamma Y_X 
\label{eq.zeta}
\end{equation}where $E_\gamma$ is the energy of the final state photon,
\begin{equation}
E_{\gamma} =\frac{1}{ 2M_{ X^{\prime} } }(M^2_{X^{\prime}}-M^2_{X}),
\end{equation} with $Y_X=n_X/n_{\gamma}^{CMB}$ being the ratio between the number density of species $X$ over the number density of CMB photons. If for each  $X^{\prime}$ particle we have the production of the pair $X+\gamma$, then $Y_{X^{\prime}}=Y_{\gamma}=Y_{X,\tau}=Y_{X,0}$, where $Y_{X,\tau}$ sets the number density of particles at a time $t=\tau$. $Y_{X,0}$ is the yield of the species $X$ today. To simplify our analytical expressions, we will assume the number density of the X is proportional to the one from dark matter. In other words, we take $n_X=f\, n_{DM}$, where $f$ is a constant factor that defines the relation between the abundance of dark matter and X particles. We include this factor to be more general, and account for the possibility of subdominant dark matter fields. One can easily rescale the total amount of energy released for any long-lived particles. Knowing that $n^{CMB}_{\gamma}= 2\zeta(2) \pi^{-2} T^3$, we get,
\begin{equation}
Y_{X}=\frac{f\, n_{DM}}{n^{CMB}_{\gamma}}= \frac{f\, \Omega_{DM} \rho_c}{M_{X} n^{CMB}_{\gamma,0}},
\label{YDMeq1}
\end{equation}which simplies to,
\begin{equation}
Y_{X} \simeq 3\cdot10^{-14}\left( \frac{\rm TeV}{M_{DM}} \right) \left( \frac{\Omega_{DM}}{0.227} \right) \left( \frac{f}{0.01}\right).
\label{Yxfinal}
\end{equation}

Using Eq.\eqref{eq.zeta} we obtain the total electromagnetic energy released,
\begin{eqnarray}
\zeta_X & = & 1.5\cdot10^{-11}\ \mbox{GeV}\times \nonumber\\
                 &   &
  \left( \frac{\Omega_{DM}}{0.227} \right) \left( \frac{f}{0.01}\right)  \left( \frac{M_{ X^{\prime} }}{M_{X}} - \frac{M_{X}}{M_{X^{\prime}} } \right).
\label{Eqzetafinal}
\end{eqnarray}

Therefore, one can use Eq.\eqref{Eqzetafinal} to determine the amount of electromagnetic energy released by a long-lived particle. Moreover, one can then pick their favorite dark sector, use Eq.\eqref{Eqzetafinal}, and map our bounds on the pair $\zeta_X-\tau$ to constrain their models. We will see further that BBN places severe constraints on dark sectors that induce electromagnetic cascades in the early universe. This reasoning will clearly show that one can explore the interplay between cosmology and particle physics and use BBN as a probe for new physics. 

Hence, in this work, we compare the results of Big Bang Nucleosynthesis (BBN) old data against the most recent Planck data, considering Planck's measurement of $\eta_{10,\textrm{CMB}}$\footnote{The subscript 10 is a reference to the fact that we are considering the values of $\eta \times 10^{10}$.}, which is in agreement with the current data \cite{ParticleDataGroup:2022pth} provided by
\begin{equation}
    \eta_{10}=\frac{n_{b}}{n_{\gamma}}=6.104 \pm 0.058.
\end{equation}

This measurement of the baryon-to-photon ratio can be used to derive a bound on the baryon relic abundance $\Omega_{b}h^{2}$ \cite{Planck:2018vyg},
\begin{equation}
    \Omega_{b}h^{2}= 0.022298\pm 0.000212,
\end{equation}
where $h$ is the reduced Hubble constant.

The BBN theory predicts the abundance of light elements like D/H, $^{3}$He,$^{4}$He and $^{7}$Li/H that can be inferred from astronomical observations. However, it is important to emphasize that these results depend on the assumed baryon-photon relationship, which can be determined by the anisotropy spectrum of the CMB or by the concordance of the abundances generated by the BBN. In Tab. \ref{Table 1} we compare the theoretical values obtained from the BBN predictions with the old observational data \cite{Cyburt:2002uv} and with the new results provided by the PDG \cite{ParticleDataGroup:2022pth}. As we can see for Y$_{P}$, D/H and $^{3}$He/H the theoretical and observational results converge, with only a small fluctuation between the values. However, when comparing the theoretical prediction with the observational data of $^{7}$Li we verify a marked divergence between the results, where $^{7}$Li/H$_{(teo)}\sim 3\times^{7}$Li/H$_{(obs)}$.  In our work, we did not derive the bounds of 7Li due to the large uncertainties. The values adopted for the abundance of the light elements are summarized in Table \ref{tablesummary}. \\
In Table \ref{tablesummary}, we show the values resulting from BBN theory, old and new observations. The latter was used to derive our bounds, as we describe below.

%%%%%%%%%%%%%%%%%%%% FIGURE 2 %%%%%%%%%%%%%%%%%%%

\begin{figure*}%[ht!]
\centering
    \subfigure[]{
    \includegraphics[width =0.9 \columnwidth]{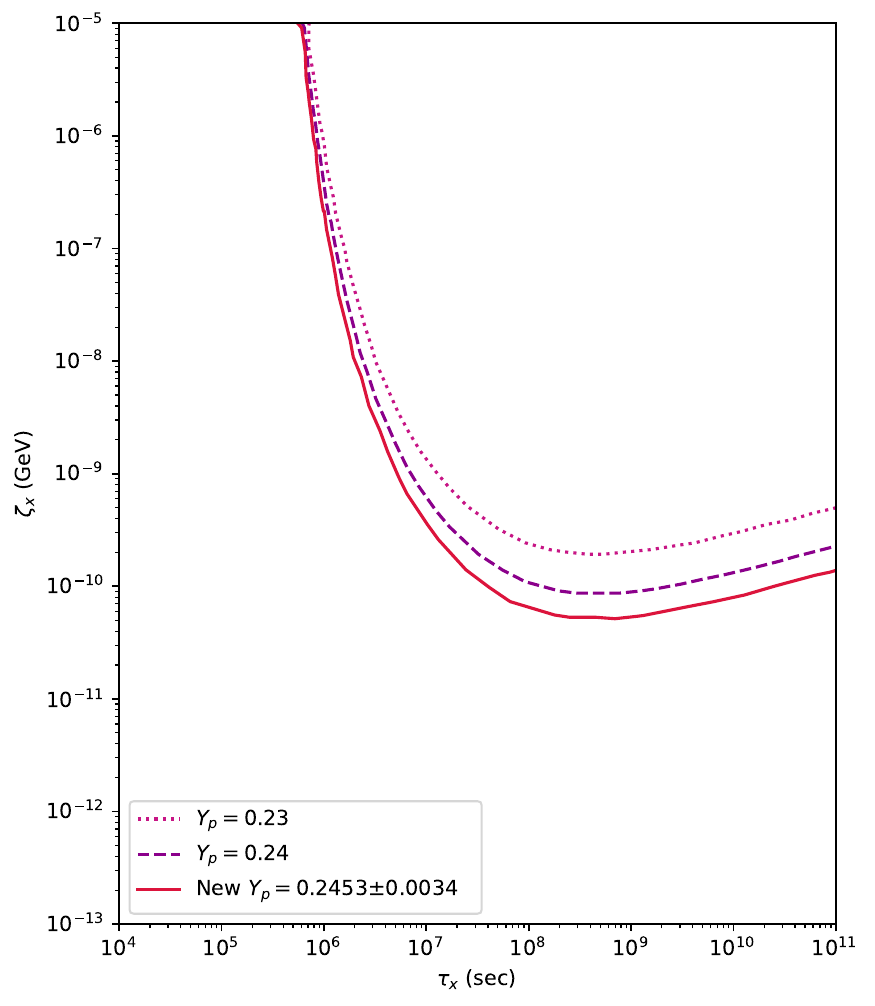}
    \label{fig:mylabela}
    }
    \subfigure[]{
    \includegraphics[width =0.9 \columnwidth]{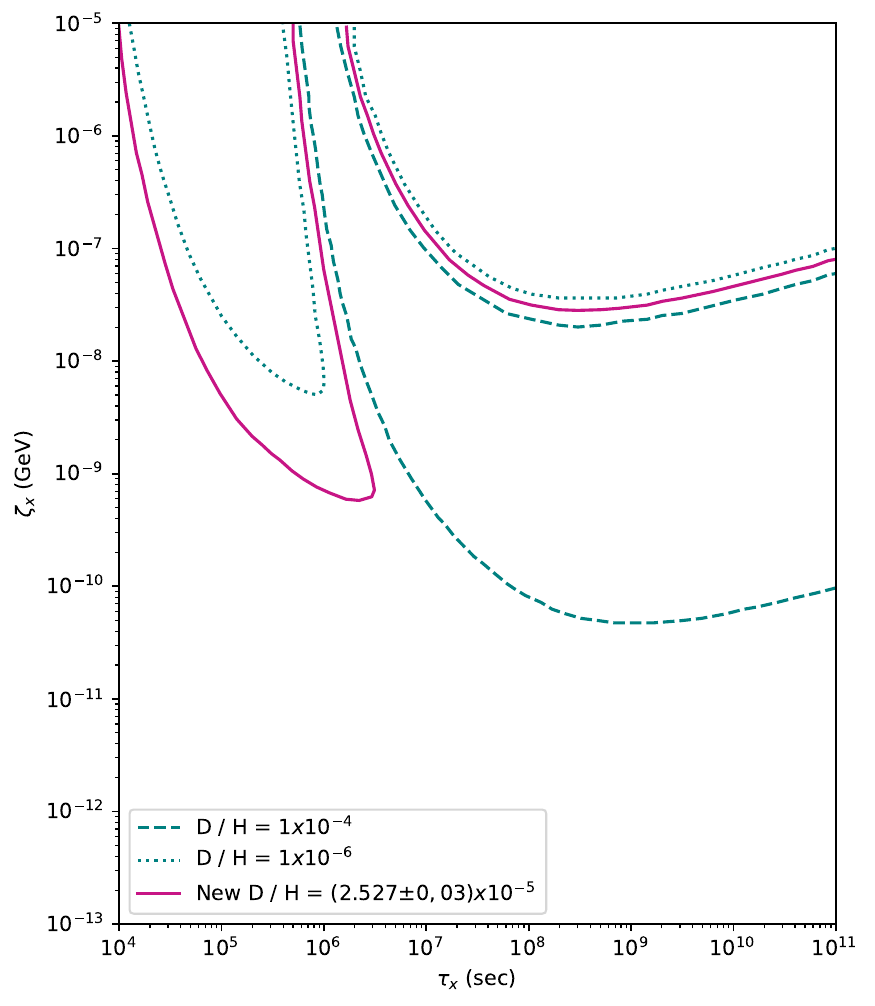}
    \label{fig:mylabelb}
    }
     \caption{Contours of the observations of the abundance of light elements in the plane $\zeta_X-\tau_X$, for $\eta_{10}=6$ considering different values for (a) mass fraction of $^{4}$He and (b) Deuterium abundance. The new contours correspond to the most recent data from the PDG \cite{Cyburt:2002uv}} 
    \label{fig:mylabel2}
\end{figure*}
%%%%%%%%%%%%%%%%%%%% FIGURE 2 %%%%%%%%%%%%%%%%%%%

%%%%%%%%%%%%%%%%%%%% TABLE 1 %%%%%%%%%%%%%%%%%%%

\begin{table*}%[h!]
\centering
\caption{Theoretical and observational analyses of the abundance of light elements in BBN.}
\label{Table 1}
\vspace{0.4cm}
\begin{tabular}{ccccccc}
\hline
Yields & Y$_{p}$ & D/H $\left(\times 10^{-5}\right)$ & $^{3}$ He/H $\left(\times 10^{-6}\right)$ & $^{7}$ Li/H $\left(\times 10^{-10}\right)$ & $^6$Li/$^7$Li & Ref. \\
\hline
Calculation & 0.248 $^{+0.001}_{-0.002}$ & 2.692 $^{+0.177}_{-0.070}$ & 9.441 $^{+0.511}_{-0.466}$ & 4.283 $^{+0.335}_{-0.292}$ &  & \\
\hline
Old Observation & 0.248 $\pm 0.001$ & 2.9 $^\pm 0.3$ & 9. $\pm 1.$ & 1.23 $\pm 0.06$ & & \cite{Cyburt:2002uv} \\
\hline
 New Observation & 0.245 $\pm 0.003$ & 2.527 $\pm 0.03$ & $\geq 11. \pm 2.$ &  1.58 $^{+0.35}_{-0.28}$ &  & \cite{ParticleDataGroup:2022pth} \\
\hline
\end{tabular}
\label{tablesummary}
\end{table*}  

%%%%%%%%%%%%%%%%%%%% TABLE 1 %%%%%%%%%%%%%%%%%%%

\section{\label{secIV}New limits for the abundance of observed light elements}

In section \ref{secIII} we saw that it is important to consider the baryon-photon number density of the early universe to study the BBN predictions. Thus, to impose stronger restrictions on the abundance of light elements, we use the results presented in section \ref{secIV} and set $\eta_{10} = 6$, varying only the lifetime between $\tau = 10^{4}$ sec and $\tau = 10^{11}$ sec. We present the new energy injection limits obtained for the new constraints of $^{4}$He and D/H, as can be seen in Fig. \ref{fig:mylabel2}.

The most probable value of the $^{4}$He abundance ($Y_{p} = 0.2453$) was found through analysis of extragalactic HII regions (ionized) \cite{ParticleDataGroup:2022pth, Aver:2020fon}. However, the recommended confidence interval for $Y_{p}$ is
\begin{eqnarray}
    0.2419 < Y_{p} < 0.2487.
\end{eqnarray}

Thus, by imposing the lower bound of $Y_{p} > 0.2419$, the constraint observed for $Y_{p}$ is
\begin{eqnarray}
    \zeta_{X}(Y_{p})< 5.2\times 10^{-11}{\rm GeV}.
\end{eqnarray} 

As $^4$He is orders of magnitude more abundant than the other elements, no significant production of $^4$He can take place. In other words, electromagnetic energy injection can only destroy $^4$He.  To understand the behavior of the curve in Fig.\ref{fig:mylabela}, we remind the reader of some fundamental aspects of electromagnetic injection. In the sudden decay approximation, where all particles decay at $t=\tau_X$, a reaction turns on when $\tau_X > 10^6 {\rm s}  \, (E_{th}/10 MeV)^2$.  Keep in mind that the photodestruction processes have an energy threshold of 20 MeV. Hence, the $^4$He destruction will be efficient only for $\tau \gtrsim 10^{6}$~s. For this reason, any small value of $\zeta_X$ will severely destroy $^4$He in contradiction to observations. One needs a very large energy injection for a shorter lifetime to alter the $^4$He abundance. This feature explains the weakening in sensitivity $\tau < 10^6$~s in Fig.\ref{fig:mylabela}. In Fig.\ref{fig:mylabela}, we also exhibit how BBN sensitivity to new physics improves with the value of $Y_p$.

Unlike $Y_{p}$, the change of the Deuterium abundance is more subtle. The D/H ratio will depend highly on the moment the energy injection took place. The most important production process is $^4$He $+\ \gamma$, because the other elements have relatively negligible abundances to produce any significant impact on the Deuterium abundance. The reactions that produce Deuterium have an energy threshold of 25 MeV, so again they are efficient only for $\tau_X  \gtrsim 10^6$~s. In other words,  Deuterium production occurs mostly when there is sizeable $^4$He destruction. If we increase $\zeta_X$ too much, the photodestruction processes are so effective that Deuterium abundance quickly decreases, leaving the universe filled with protons only. For $\tau< 10^6$~s, the Deuterium production does not happen, and only destruction occurs. Eventually, for $\tau < 10^3$~s, the Deuterium destruction freezes. The competing processes, around $10^6$~s, explain the shape of the curve in Fig.\ref{fig:mylabelb}. In Fig.\ref{fig:mylabelb}, we plot an old limit that delimited the Deuterium abundance to lie between $10^{-4}<$ D/H $< 10^{-6}$  and show ours which is based on a much more restrictive statistical range for D/H, namely \cite{ParticleDataGroup:2022pth, Cooke:2017cwo},

\begin{eqnarray}
    2.227\times 10^{-5} < \left(\frac{{\rm D}}{{\rm H}}\right)_{p} < 2.827\times 10^{-5}.
\end{eqnarray}

As we do not allow a large abundance for Deuterium, i.e. a high D/H, our limit weakens for $\tau> 10^7$~s. From Fig.\ref{fig:mylabelb} we concluded that,

\begin{eqnarray}
    \zeta_{X}({\rm D/H})\leq 6.0\times 10^{-10}{\rm GeV}, {\rm for}\, \tau \sim 10^6~s,
\end{eqnarray}
\begin{eqnarray}
    \zeta_{X}({\rm D/H}) \leq 3\times 10^{-8}{\rm GeV},  {\rm for}\, \tau \sim 10^8-10^{10}~s.
\end{eqnarray}

\begin{figure}[h!t]
    \centering
    \includegraphics[width=\columnwidth]{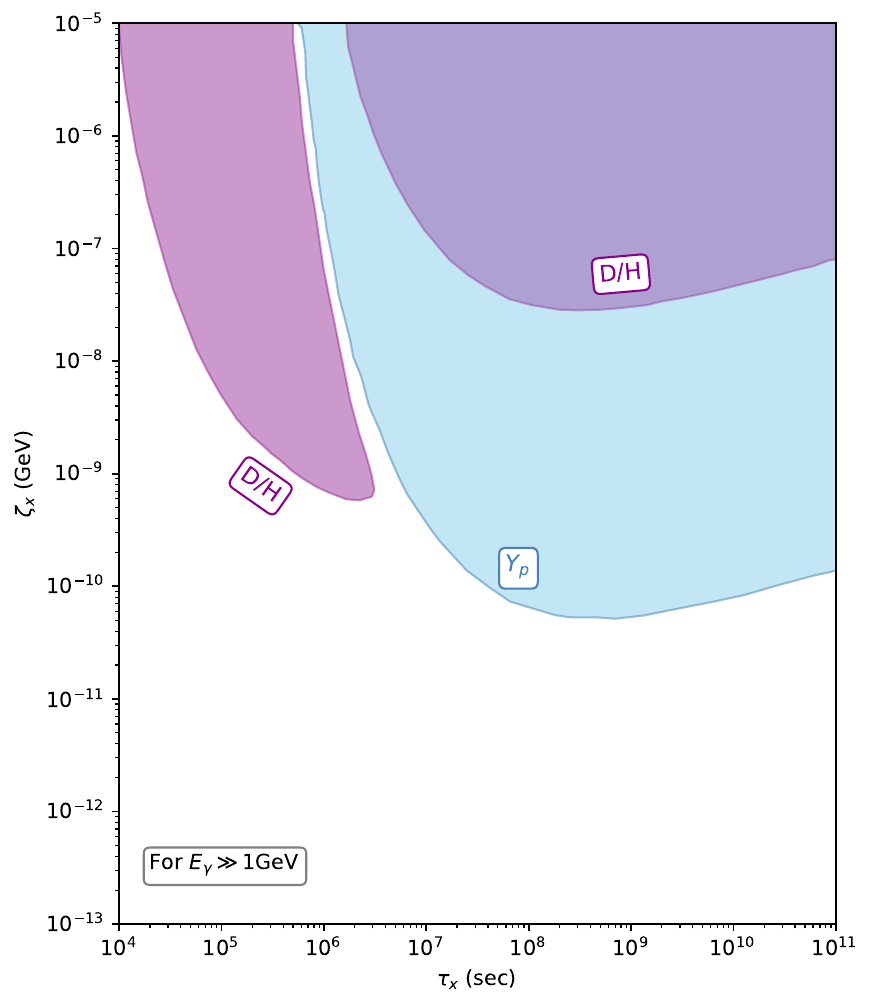}
    \caption{Summary of the upper limits on the electromagnetic energy injection, $\zeta_X$, as a function of the lifetime of the decaying particle. It is clear that Deuterium and $^4$He which have reliable primordial abundances suffice to probe electromagnetic energy injection episodes that took place from $10^2$~s to $10^{11}$~s.}
    \label{fig:mylabel3}
\end{figure}

We combine all upper limits into Fig. \ref{fig:mylabel3}, so one can clearly see the importance of each element to constraining electromagnetic injections in the early universe. The shaded regions delimit the exclusion limits from the D/H abundance (purple). The blue region comes from the $^4$He abundance. Deuterium and $^4$He which have reliable primordial abundances suffice to probe electromagnetic energy injection episodes.

We have assumed sufficiently large injection energies, where the most rapid interactions in this cascade are pair production $\gamma \gamma_{CMB} \rightarrow e^+e^-$ off of background photons and inverse Compton scattering processes. In this case, these processes quickly redistribute the injected energy, and the nonthermal photon spectrum reaches a quasi-static equilibrium, i.e. a universal spectrum. Hence, we need $E_\gamma > E_C= m_e^2/(22 T)$.  In our work, we are considering the decaying scenario, $X^\prime \rightarrow X+\gamma$, with $M_{X^\prime} \gg M_X$, thus $E_\gamma \sim M_{X^\prime}/2$. Therefore, we assume that $M_X \gg  E_C$ at any given temperature or lifetime in agreement with \cite{Forestell:2018txr}. Approximately that refers to energy injections larger than $1$~GeV. For energy injections below 1 GeV, the computation of BBN bounds are non-trivial, and we leave this for future work. Recent developments in that regard have been done in \cite{Forestell:2018txr, Depta:2021}.

%%%%%%%%%%%%%%%%%%%% FIGURE 3 %%%%%%%%%%%%%%%%%%%
%%%%%%%%%%%%%%%%%%%% FIGURE 3 %%%%%%%%%%%%%%%%%%%

\section{\label{secV}Discussion and Conclusions}

In this article, we have revisited the impact of electromagnetic injection episodes on the abundance of light elements, such as $^4$He, and D. We were particularly interested in the decaying scenario, $X^\prime \rightarrow X +\gamma$, where $X$ had an abundance proportional to the dark matter one. We have computed the change in the abundance of those elements as a function of the total electromagnetic energy released and the lifetime of the decaying particle, $X^\prime$. In our analysis, we adopted the primordial abundance of the light elements reported from recent observations and used the baryon-photon ratio indicated by Planck data, $\eta^{cmb}_\gamma = 6.104 \times 10^{-10}$.  Firstly, we derived separately the bounds from $^4$He and $D$ abundances and later combined them all into one figure. We concluded that Deuterium and $^4$He which have reliable primordial abundances can already rule out energy injections down to $\zeta_X \sim 10^{-9}$~GeV for $\tau \sim 10^5-10^6$~s, and $\zeta_X \sim 10^{-10}$~GeV for $\tau \sim 10^8-10^{11}$~s.

Anyway, we point out that the scenario covered in our study appears in supersymmetric models, extended gauge sectors, and decaying dark matter models. Hence, it is clear that BBN represents a great laboratory for well-motivated dark sectors that induce energy injection in the early universe. 

\acknowledgments

The authors thank Jacinto Paulo and Alvaro de Jesus for their discussions. DRS thanks for the support from CNPQ under grant 303699/2023-0. FSQ is supported by Simons Foundation (Award Number:1023171-RC), FAPESP Grant 2018/25225-9, 2021/01089-1, 2023/01197-4, ICTP-SAIFR FAPESP Grants 2021/14335-0, CNPq Grants 307130/2021-5, and ANID-Millennium Science Initiative Program ICN2019\textunderscore044. Y.S.V. expresses gratitude to São Paulo Research Foundation (FAPESP) under Grant No. 2018/25225-9 and 2023/01197-4. JRA acknowlodges the support from Coordenação de Aperfeiçoamento de Pessoal de Nível Superior (CAPES) under grant No. 88887.827362/2023-00. L.G. thanks the support from Coordenação de Aperfeiçoamento de Pessoal de Nível Superior (CAPES) under grant No. 88887.704425/2022-00. RMPN is thankful to CNPQ for the support under grant 151428/2022-0. LA acknowledges the support from Coordenação de Aperfeiçoamento de Pessoal de Nível Superior (CAPES) under grant 88887.827404/2023-00.
R. Silva acknowledges financial support from CNPq (Grant no. 307620/2019-0).

\def\bibsection{\section*{References}}

\bibliographystyle{JHEPfixed.bst}
\bibliography{references.bib}

\end{document}